\documentclass[aps,pra,twocolumn,showpacs,amsmath,amssymb,groupedaddress,floatfix]{revtex4-1}

\usepackage{graphicx}
\usepackage{dcolumn}
\usepackage{bm}
\usepackage{amssymb}
\usepackage{multirow}
\usepackage{SIunits}
\usepackage{float}
\usepackage{color}

\begin{document}

\title{Capture Dynamics of Ultracold Atoms in the Presence of an Impurity Ion}

\author{J. M. Schurer}
    \email{jschurer@physnet.uni-hamburg.de}
\author{A. Negretti}
\author{P. Schmelcher}

\affiliation{Zentrum f\"ur Optische Quantentechnologien, Universit\"at Hamburg, Luruper Chaussee 149,
22761 Hamburg, Germany}
\affiliation{The Hamburg Centre for Ultrafast Imaging, Universit\"at Hamburg, Luruper Chaussee 149, 22761
Hamburg, Germany}

\date{\today}


\begin{abstract}
We explore the quantum dynamics of a one-dimensional trapped ultracold
ensemble of bosonic atoms triggered by the sudden creation of a single
ion. The numerical simulations are performed by means of the ab initio
multiconfiguration time-dependent Hartree method for bosons which
takes into account all correlations. The dynamics is analyzed via a cluster
expansion approach, adapted to bosonic systems of fixed particle number,
which provides a comprehensive understanding of the occurring many-body
processes. After a transient during which the atomic ensemble
separates into fractions which are unbound and bound with
respect to the ion, we observe an oscillation in the atomic density which we attribute to the additional
length and energy scale induced by the attractive long-range atom-ion interaction.
This oscillation is shown to be the main source of spatial coherence
and population transfer between the bound and the unbound atomic fraction.
Moreover, the dynamics exhibits collapse and revival behavior caused by
the dynamical build-up of two-particle correlations demonstrating that
a beyond mean-field description is indispensable.
\end{abstract}

\maketitle

\newcommand{\comm}[1] { {\textcolor{red}{(* #1 *)}} }   
 \newcommand{\new}[1]{{\textcolor{red}{#1}}} 
 \newcommand{\old}[1]{{\textcolor{Gray}{#1}}} 

\newcommand{\mb}[1]{\mathbf{#1}}						
\newcommand{\mt}[1]{\mathrm{#1}}						
\renewcommand{\d}{\mt{d}}							

\newcommand{\Hut}[1]{\hat{#1}}

\newcommand{\zi}{\mathrm{i}}
\newcommand{\Komma}[0]{,\,}

 \newcommand{\vecB} [1] {\mb{#1} }
 \renewcommand{\vec} [1] { {\vecB{#1}} }

\renewcommand{\Re}{ \mathrm{Re} }                                               
\renewcommand{\Im}{ \mathrm{Im} }                                               
    \definecolor{background}{gray}{0.8}
\newcommand{\gdag}           { {\textcolor{background}{\dagger}} }   
\newcommand{\wdag}           { {\textcolor{white}{\dagger}} }     


\newcommand{\bra}[1]{\langle #1|}
\newcommand{\ket}[1]{|#1\rangle}
\newcommand{\ketO}[1]{#1\rangle}

\newcommand   {\<}           { \langle }
\renewcommand {\>}           { \rangle }

\newcommand{\ad}        [1] { \Hut{a} ^\dagger _{ #1 } }
\renewcommand{\a}       [1] { \Hut{a} ^\wdag   _{ #1 } }
\newcommand{\ag}        [1] { \Hut{a} ^\gdag   _{ #1 } }

\newcommand{\psiOd}        [1] { \Hut{\Psi} ^\dagger {\left( #1 \right)} }
\newcommand{\psiO}       [1] { \Hut{\Psi} ^\wdag   { \left( #1 \right) } }
\newcommand{\psiOg}        [1] { \Hut{\Psi} ^\gdag   { \left( #1 \right) } }

\newcommand{\ald}        [1] { \Hut{\alpha} ^\dagger _{ #1 } }
\newcommand{\al}       [1] { \Hut{\alpha} ^\wdag   _{ #1 } }
\newcommand{\alg}        [1] { \Hut{\alpha} ^\gdag   _{ #1 } }

\section{Introduction}\label{sec:intro}

The theoretical description of degenerate atomic quantum gases has advanced significantly in the past two 
decades. It relies particularly on the separation of length scales of the external 
trapping potential, the inter-particle distances and the range of atomic interactions. In the ultracold 
regime, this allows to approximate the interaction by a 
contact potential (or simply pseudopotential) \cite{Huang1957a}, which has proven tremendously powerful 
when applied to bosonic gases. The latter holds especially for the mean-field 
approximation leading to the well-known Gross-Pitaevsikii (GP) equation \cite{Gross1963,Pitaevskii1961} which 
describes the condensed state of a weakly interacting bosonic 
gas yielding a variety of phenomena such as collective excitations, solitons, and vortices 
\cite{Pitaevskii2003,Pethick}. For state-of-the-art methods, such as the density-matrix 
renormalization group \cite{Schollwock2011} and the multiconfiguration time-dependent Hartree method 
for bosons (MCTDHB) \cite{Alon2008} as well as its multi-layer extension (ML-MCTDHB) 
\cite{Cao2013,Kronke2013}, the 
pseudopotential represents an essential simplification for the 
quantum dynamical description of many-boson systems in order to investigate the physics beyond the 
mean-field approximation.
Even though the short-range interaction has proven to lead to exotic 
states of quantum matter like supersolidity \cite{Buchler2003} or crystalline phases 
\cite{Buchler2011} and therefore represents an important case, not all ultracold atomic systems exhibit these 
short-range interactions. An example for the latter are chromium atoms which possess long-range magnetic 
dipolar interactions \cite{Stuhler2005,Lahaye2009}.

Hybrid atom-ion systems represent a specific class of systems with a new scale of interactions due to the 
interplay between the charge of an ion and the induced dipole moment of a neutral atom
\cite{Seaton1977}. They have recently become available experimentally
\cite{Smith2005,Zipkes2010,Schmid2010a,Harter2012a} and attracted increasing interest. 
Most of the current experiments, however, are based on the Paul trap scheme, whose drawback is the 
so-called micromotion which so far prevents from reaching the ultracold regime as shown in theoretical 
classical and quantum analyses \cite{Cetina2012,Krych2015}. Corresponding studies showed that a large 
ion-atom mass 
ratio might help circumventing this limitation \cite{Cetina2012,Joger2014,Krych2015}, although it 
is not yet clear whether the 
s-wave regime can be reached. Still, a recent detailed study has shown 
that specifically for the atom-ion pair $\mt{Li}$-$\mt{Yb}^{+}$ the ultracold regime can be reached 
experimentally \cite{Tomza2015}.
Given these findings, it is desirable to extend the theoretical understanding of ultracold neutral
quantum many-body systems by the presence of ions. This attractive interaction 
induces, additional to the confinement, a further length and energy scale and  it is therefore expected to 
lead to intriguing effects such as the 
formation of molecular ions \cite{Cote2002} and ion-induced density bubbles in the atomic cloud 
\cite{Goold2010}. 
Apart from this fundamental point of view, these hybrid systems show versatile applications in quantum 
information processing as, for example, the controlled creation of entanglement 
\cite{Joger2014,Gerritsma2012}, the realization of quantum gates \cite{Doerk2010}, and the simulations of 
solid-state systems \cite{Bissbort2013}.

In a previous study, we have investigated in detail the ground-state properties of an atom-ion hybrid system, 
consisting of a single static ion in the center of a bosonic atomic cloud. The dependence of relevant 
observables on the atom number 
and the interaction strength has been analyzed and we showed that the 
presence of an ion strongly affects them \cite{Schurer2014}.
For weakly interacting atoms, we found that the ion impedes the transition to the Thomas-Fermi regime while 
for strong atom-atom interaction it modifies the fragmentation behavior depending on the atom number 
parity.	
In the present work, we explore the impact of the second length and energy scale generated 
by the atom-ion interaction onto the dynamics of the atomic cloud.
In particular, we envision the scenario in which we create a single ionic impurity in an 
ensemble of $N$ interacting atomic bosons. If the atomic cloud, initially prepared in the 
ground state of a harmonic trap, and the ion are suddenly brought into contact, this process resembles, to 
some extent, the sudden ionization of a single impurity atom within the atomic cloud. Here, however, we still 
neglect the ionic motion which is justified in case the ion is tightly trapped.
We will see in the following that the second length and energy scale induces a coherent oscillation between states bound 
in the atom-ion potential and states of the harmonic trap. This oscillation occurs in addition to the usual 
harmonic excitation and reveals a collapse and revival behavior caused by the dynamical build-up of 
correlations.

This work is organized as follows. In Sec. \ref{sec:Model}, we define the setup and explain the model of 
our hybrid atom-ion system. In addition, we introduce a cluster-expansion scheme which enables us to 
distinguish the single-particle dominated physics from the processes induced by the build-up of (quantum) 
correlations.
In Sec. \ref{sec:DynEvol}, we explore the dynamical evolution of the hybrid system and identify the most
important excited modes by means of our cluster-expansion approach. Sec. 
\ref{sec:conver} contains a brief convergence 
analysis which shows the necessity to use an advanced method like MCTDHB.
Finally, we summarize our findings in Sec. 
\ref{sec:concl} together with  the conclusions and an outlook on future investigations.

\section{Model and Theoretical Approach}\label{sec:Model}

In the following, we introduce our model and define the process to 
initiate the dynamics. Further, we briefly outline the MCTDHB used for the simulations, define important 
quantities for the analysis, and introduce our cluster expansion approach which we exploit in order to 
analyze the many-body wave function.

\subsection{Model of the System}\label{subsec:ModelofSys}

We consider $N$ interacting bosonic atoms in a one-dimensional (1D) harmonic trap at zero temperature 
initially prepared in the ground state of the system which we compute by imaginary time propagation 
\cite{Kosloff1986} of an initial guess wave function. The 
short-range intra-atomic interaction is modeled by a contact pseudopotential. Into this atomic 
cloud, we immerse a single trapped impurity atom which does not interact with the other atoms. This could 
be achieved by tuning the inter-atomic interaction to zero by exploiting a Feshbach resonance \cite{Chin2010}. 
At time $t_0 = 0$ this impurity atom is ionized by a laser pulse such that a single ion is created 
within the atomic cloud. The ionizing laser pulse is assumed to be far detuned from any possible resonance 
of the atomic cloud. Moreover, we assume pulse durations such 
that the ionization process takes place on much faster time scales than 
any possible response of the atomic cloud. This allows us to treat the ionization as an effectively 
instantaneous process.
If the atomic impurity is trapped in a sufficiently deep and tight trap, the proposed scenario 
can be indeed achieved experimentally, as recently reported for optical trapping of a single ion 
\cite{Huber2014}.
Thus, at time $t_0 = 0$, the ion is assumed to be statically trapped at $z_\mt{I}=0$ in the atomic cloud. The 
motional excitations following this ionization process result in a rich dynamics which we analyze in Sec. 
\ref{sec:DynEvol}.

The interaction between the ion at $z_\mt{I}$ and an atom at 
$z_\mt{A}$ behaves in 1D at large distances as $-\alpha e^2/(2( z_\mt{A} - z_\mt{I})^4)$ 
 up to a minimal cutoff distance $R_\mt{1D}$ \cite{Idziaszek2007}, where $\alpha$ is the polarizability of 
the atoms and $e$ the elementary charge. 
For numerical many-body simulations with MCTDHB it is more convenient, however, to define a model 
potential as \cite{Schurer2014}:
\begin{equation}
 \label{eq:modelPotpaper}
 V_{\mt{mod}}(z) = \mt{v}_0 e^{-\gamma z^2} - \frac{1}{z^4 + 1/\omega}.
\end{equation}
Here $z$ denotes the relative coordinate $ z_\mt{A} - z_\mt{I}$.
The model parameters $\mt{v}_0$,$\gamma$, and $\omega$ are determined by the short-range quantum defect 
parameters. The above potential asymptotically approach the $-1/z^4$ behavior at large distances, whereas at 
short distances it has a barrier which is designed such that the quantum defect theory results  
for the atom-ion scattering \cite{Idziaszek2009} can be reproduced.
Additionally to the harmonic confinement, this interaction introduces a second length $R^* =\sqrt{\alpha e^2 
m/\hbar^2} $ and energy scale $E^* = \hbar^2/(2m {R^*}^2)$ to the system, where $m$ is the mass of a single 
neutral atom.

\begin{figure}
 \centering
 \includegraphics[width=\linewidth]{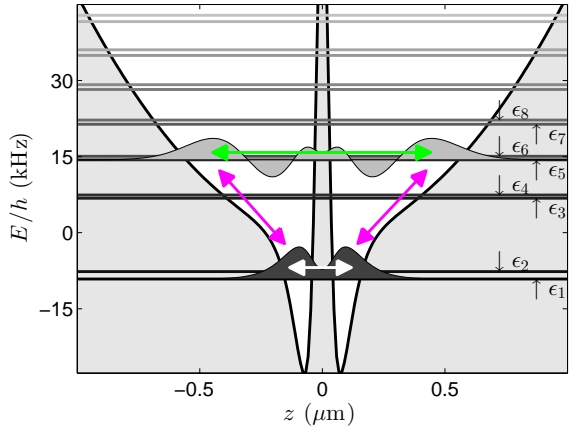}
 \caption{ The effective potential (gray shaded area in background) consisting of the harmonic trap and 
the atom-ion interaction potential for the $^{87}\mt{Rb}$ atom. Further, we show the single-particle energy 
levels $\epsilon_j$ indicated by straight horizontal lines. In addition, the energetically lowest 
single-particle state $\phi^0_1(z)$ (dark gray area) and the trap state $\phi^0_5(z)$ (light gray area) are 
sketched with arbitrary but equal scaling. Possible types of dynamics occurring in the system are indicated by 
arrows.}
\label{fig:potLevels}
\end{figure}

In summary, the system can be described by the following many-body Hamiltonian (in $E^*$ and $R^*$ units)
\begin{align}\label{eq:Hamiltonian}
  \hat{H} =& \sum_{i=1}^N \underbrace{\left[ - \frac{\partial^2}{\partial z_i^2} + \frac{1}{l^4}z_i^2  
+ \theta(t-t_0)  V_{\mt{mod}}(z_i) \right]}_{\hat{h}_i} \nonumber \\
&+  g\sum_{i<j}^N  \delta(z_i-z_j) 
\end{align}
with the harmonic trap of frequency $\omega_0$ and characteristic length $l=\sqrt{\hbar/(m\omega_0)}/R^*$, 
and the intra-atomic contact interaction strength $g$. The ionic part is switched on at time $t_0$ by a 
step-function $\theta(t-t_0)$. 
Thereby, we term the Hamiltonian of a single boson as $\hat{h}_i$.
In the following, we denote the stationary eigenenergies of $\hat{h}_i (t>t_0)$ for fixed $t$ with 
$\epsilon_j$ and
the corresponding single-particle eigenstates with $\phi^0_j(z_i)$ such that 
$\hat{h}_i(t>t_0) \phi^0_j(z_i) = \epsilon_j \phi^0_j(z_i)$.
Hereafter, for the sake of numerical convenience, we set $l=0.5 R^*$, which would
corresponds to $\omega_0= 2\pi \cdot 3.3\,\kilo\hertz$ and $l = 188\, \nano\meter$  for 
$^{87}$Rb atoms. We will see that this choice does not affect qualitatively the observed dynamics. 
Furthermore, we choose a 
weak intra-atomic interaction strength $g=2 E^* R^*$ 
and small atomic ensembles consisting of $N=2$ up to 
$N=10$ neutral atoms. Besides, we fix the model parameters to $ \omega = 80 (R^*)^{-4}$, $\mt{v}_0 = 3\omega$
and $\gamma = 4\sqrt{10\omega}$. We refer here to Ref. \cite{Schurer2014} for a detailed discussion of the 
chosen parameters as well as for the experimental conditions needed for the quasi-1D regime.
Our above choice leads to two bound states for the atoms in the atom-ion potential which are localized on 
both sides of the ion but vanish at $z_\mt{I}$. Even though we neglect the motion of the ion, we term the two 
states below $E=0$ \textit{bound states} while we refer to the remaining states as \textit{trap states} 
(with $E>\hbar\omega_0/2$).
In Fig. \ref{fig:potLevels}, we show the total potential for the atoms together with the energies $\epsilon_j$ 
as well as the lowest single-particle state $\phi^0_1(z)$, bound in the atom-ion potential, and the state 
$\phi^0_5(z)$. The arrows indicate the possible processes that can occur: within the trap (green arrow), in 
the atom-ion potential (white arrow), and between those two scales (oblique magenta arrows).

\subsection{Theoretical Approach}

We explore the quantum dynamics of the many-body system described by the wave function $|\Psi\rangle$ 
by means of the numerically exact ab initio method MCTDHB. Its main idea is that by using $m$ 
\textit{time-dependent} variationally optimized single-particle basis functions, the 
number of basis functions can be kept rather small.
More precisely, the many-body wave function $|\Psi\rangle$ for $N$ bosons is expanded in bosonic number 
states $|\vec{n}(t)\rangle$
\begin{equation}\label{eq:ansatz}
 |\Psi(t)\rangle = \sum_{\vec{n}|N} A_{\vec{n}}(t) |\vec{n}(t)\rangle
\end{equation}
in order to take into account the indistinguishability of the bosons.
Note that in a number state $|\vec{n}(t)\rangle$, each boson occupies one of the $m$ time-dependent 
single-particle functions (SPFs) $|\Psi_j(t)\rangle$ and that the vector $\vec{n} = (n_1,\cdots,n_{m})$ 
contains the occupation numbers $n_j$ of every SPF. Besides, the sum in Eq. \eqref{eq:ansatz} goes over all 
possible $\vec{n}$ with $\sum_j n_j = N$, which is denoted by the symbol $\vec{n}|N$.
With this ansatz for the many-body wave function, the temporal evolution of the wave function 
$|\Psi(t)\rangle$ is obtained by means of the Dirac-Frenkel variational principle \cite{Dirac1930,Frenkel1934} 
which guarantees a variational optimal many-body solution. We would like to emphasize that not 
only the coefficients $A_{\vec{n}}(t)$ but also the SPFs $|\Psi_j(t)\rangle$ are adapted in 
time to the many-body dynamics in order to allow for the largest possible overlap between the ansatz 
\eqref{eq:ansatz} and the true many-body wave function.
We refer for a detailed description of the method to Refs. \cite{Alon2008,Cao2013,Kronke2013}.

The analysis of the full many-body wave function $\ket{\Psi}$ is generally a complicated task due to the 
underlying high dimensionality 
resulting from the $N$ degrees of freedom. In order to analyze the time-dependent many-body 
wave function in detail, we inspect two different quantities: The one- and two-particle reduced 
density matrices which allow for the investigation of spatial coherence and correlations \cite{Sakmann2008}, 
and \textit{clusters} enabling us to analyze coherence and correlations in 
terms of any single-particle basis. We transfer and adapt the notion of clusters from Ref. 
\cite{Kira2006} to bosonic systems of few particles.

\subsubsection{Density Matrices}

The reduced one- and two-particle density matrices are defined via the expectation values of the field 
operators $\psiOd{x}$ and $\psiO{x}$ as $\rho_1(x,y,t) = \< \psiOd{x,t} \psiO{y,t} \> $ and 
$\rho_2(x,y,y',x',t) = \< \psiOd{x,t} \psiOd{y,t} \psiO{y',t} \psiO{x',t} \>$, respectively.
Their spectral decomposition can be written in terms of the natural populations $\lambda_j(t)$ and the natural 
orbitals $\Phi_j(x,t)$ 
\begin{equation} \label{eq:natorbs}
 \rho(x,x',t) = \sum_j \lambda_j(t) \Phi_j^*(x,t)\Phi_j(x',t),
\end{equation}
and the natural populations $\gamma_j(t)$ and natural geminals $\Phi_j(x,x',t)$
\begin{equation}\label{eq:natgems}
 \rho_2(x,y,y',x',t) = \sum_j \gamma_j(t) \Phi_j^*(x,y,t) \Phi_j(x',y',t),
\end{equation}
respectively.
This natural representation of the reduced density matrices has several advantages. For example, the natural 
populations can be used to identify the degree of fragmentation of the system \cite{Penrose1956} and to 
judge the convergence of our numerical simulations \cite{Beck2000}. Further, the natural orbitals build always 
a very suitable single-particle basis set for the description of the dynamical system, hence in many cases 
only  a few basis functions are needed to represent the (many-body) wave function in this basis. Despite 
these advantages, the analysis and a deep physical understanding of the quantum dynamics 
in this basis is very difficult, since the natural orbitals are 
time-dependent eigenfunctions of the density matrix in its spatial representation. Given this, we use instead 
the so-called clusters for the detailed analysis of the many-body dynamics. We define in the 
following the notion of clusters in the framework of the cluster-expansion approach.

\subsubsection{Definition of Clusters and Correlations}\label{sec:defCluster}

The cluster-expansion approach is a very powerful technique to describe the quantum dynamics of interacting 
many-body systems because it allows for a systematic truncation of the Bogolyubov-Born-Green-Kirkwood-Yvon 
(BBGKY) hierarchy \cite{Wyld1963}. By expanding $M$-particle expectation values into cumulants 
or correlated clusters, a consistent theory up to the single-, two-, or even $M$-particle level can be 
developed \cite{Purvis1982}. In this spirit, we shall use the cluster-expansion approach in order to separate  
the many-body dynamics, obtained by means of MCTDHB, into single- and two-particle contributions 
which will enable us to get more physical insight. In particular, it will be useful for the identification  
and classification of the most relevant excitations of the system that are created during the dynamical 
evolution (see Sec. \ref{sec:DynEvol}). 

To this end, we will first briefly review the cluster-expansion approach, which will be helpful for a better 
understanding of our modifications  to the traditional approach. Indeed, we shall adapt 
the traditional cluster-expansion to bosonic systems with a fixed particle number.
This constraint is particularly important for atomic ensembles of only a  few particles.

To begin with, let us consider an arbitrary orthonormal time-independent single-particle basis 
$\phi_j(x)$  and the annihilation and creation operators $\ad{j}$ and $\a{j}$ for the corresponding 
single-particle state, respectively.
Then, the expectation value of any observable can be expressed in terms of the $M$-particle 
expectation values or $M$\textit{-particle clusters} 
\begin{equation}
 \bra{\Psi} M_{\vec{j},\vec{k}} \ket{\Psi} = \langle M_{\vec{j},\vec{k}} \rangle = \langle \prod_i^M \ad{j_i} 
\prod_i^M \a{k_i} \rangle
\end{equation}
with the index sets $\vec{j} = \{j_1,...,j_M \}$ and $ \vec{k} = \{ k_1,...,k_M\}$ and $M \leq N$. These are 
nothing else but the matrix elements of the $M$-particle reduced density matrix in an arbitrary basis and 
can be, most conveniently for MCTDHB, obtained from the spectral representation of the $M$-particle 
reduced density matrix, as shown for 
$M=1$ and $M=2$ in App. \ref{sec:deriveClusters} . 

Now, the single-particle properties of the system are given by the single-particle 
clusters $\< \ad{i}  \a{j} \>$, also called \textit{singlets}. We distinguish between 
\textit{occupations} $f_j = \langle \ad{j} \a{j} \rangle$, which describe the population of the state 
$\phi_j(x)$, and \textit{coherences} $p_{ij} = \langle \ad{i} \a{j}\rangle$ ($i\neq j$) which can be 
understood as a transition amplitude between the $i$-th and the $j$-th state.

Starting from the singlets, the cluster-expansion is recursively build-up based on the consistent 
factorization of an $M$-particle cluster into independent particles (singlets), correlated pairs, correlated 
three particle clusters, up to correlated $M$-particle clusters \cite{Kira2006,Kira2012}, that is:
\begin{align}
  \langle 1 \rangle &= [\langle 1 \rangle]_\mt{S} \\
  \langle 2 \rangle &= [\langle 2 \rangle]_\mt{S} + \Delta \langle 2 \rangle_\mt{C} \\
  \langle 3 \rangle &= [\langle 3 \rangle]_\mt{S} + [\Delta \langle 2 \rangle_\mt{C} \langle 1 
\rangle]_\mt{D}   + \Delta \langle 3 \rangle_\mt{C} \\
  &\vdots \nonumber
\end{align}
Here the terms $[\langle M \rangle]_\mt{S}$ represent the single-particle contributions,  while 
the terms $\Delta \langle M \rangle_\mt{C}$ contain the correlated part of the $M$-particle cluster. 
Note that we omitted the indices for brevity such that all cluster products, denoted in square 
brackets, include a sum over all unique permutations. For instance, for 
the two-particle clusters $\< \ad{k}  \ad{q} \a{q'} \a{k'} \>$, the single-particle contributions are defined 
as $ [\langle 2 \rangle]_\mt{S} \equiv [\< \ad{k}  \ad{q} \a{q'} \a{k'} \>]_\mt{S} :=  \< \ad{k}  \a{q'} \> 
\< \ad{q}  \a{k'} \> + \< \ad{k} \a{k'}\>\< \ad{q}  \a{q'} \>$. Given this, the correlated part of the 
two-particle cluster is 
given by $\Delta \langle 2 \rangle_\mt{C} := \langle 2 \rangle  - [\langle 2 
\rangle]_\mt{S}$. Now the correlated clusters $\Delta \langle M \rangle_\mt{C}$ with $M>2$ can be determined 
recursively which completes the formulation of the cluster expansion.

At this point a remark is in order: For bosonic systems with a fixed number of particles (i.e., in a 
number-conserving theory), even in a GP
type mean-field state (i.e., with only one single-particle orbital), the term $\Delta \langle M 
\rangle_\mt{C}$ is non-zero because of the bosonic symmetry.
This implies that the systematic truncation of the BBGKY hierarchy, for which the correlated parts of any 
$M$-particle cluster beyond a certain size ($M>M_\mt{T}$) have to be neglected, cannot be performed in this 
case. Note that this issue does not arise neither for fermions nor for bosons with particle number 
fluctuations. Thus, in order to circumvent this problem, we shall introduce a slightly 
different definition of the correlated parts $\Delta \langle M \rangle_\mt{C}$.
Our strategy will be to define the correlated parts of any $M$-particle cluster in such a way that all the 
$(M-1)$-particle contributions are indeed removed. As a result of such a strategy, for example, the 
correlated parts automatically vanish in any mean-field state. 

To this end, let us note that an $M$-particle cluster contains all the information about the 
$M'$-particle clusters with $M'<M$. This can be easily seen by using the recursive relation between the 
density matrices
\begin{align}
& \rho_{M-1}(x_1,\cdots,x_{M-1},x_{M-1}',\cdots,x_1') =  \nonumber \\ 
& \frac{1}{N-M+1} \int \d{x_M} \rho_{M}(x_1,\cdots,x_M,x_M,x_{M-1}',\cdots,x_1') 
\end{align}
for $M\ge2$, which leads to 
\begin{equation}
 \langle (M-1)_{\vec{j},\vec{k}} \rangle = \frac{1}{N-M+1}\sum_q \langle M_{\{\vec{j},q\},\{q,\vec{k}\}} 
\rangle.
\end{equation}
In order to identify the correlated part of an $M$-particle cluster, we decompose the $M$-particle cluster 
into two parts: one consisting of all contributions  from clusters with $M'<M$ , denoted by $\langle 
M_{\vec{j},\vec{k}} \rangle_{<M}$, and one which contains only the $M$-particle contributions, that is,
\begin{equation}
  \langle M_{\vec{j},\vec{k}} \rangle  :=  \langle M_{\vec{j},\vec{k}} \rangle_{<M} +  \Delta \langle 
M_{\vec{j},\vec{k}} \rangle. \label{eq:separation}
\end{equation}
There are several ways one could perform such a decomposition. For instance, in the traditional 
cluster-expansion approach, as we have discussed above, one would choose the following definition for the 
single-particle part of the two-particle cluster:
$\< 2_{\{k,q \}\{q',k' \}} \>_{<2} =  \< \ad{k}  \a{q'} \> \< \ad{q}  \a{k'} \> + \< \ad{k} \a{k'}\>\< 
\ad{q}  \a{q'} \>$. Here, however, we shall define the term $\langle 
M_{\vec{j},\vec{k}} \rangle_{<M}$ by requiring that it has to  fulfill the condition
\begin{equation}
  \langle (M-1)_{\vec{j},\vec{k}} \rangle = \frac{1}{N-M+1}\sum_q \langle M_{\{\vec{j},q\},\{q,\vec{k}\}} 
\rangle_{<M}. \label{eq:hierCons}
\end{equation}
This expression is assumed to hold for any many-body quantum state and implies that $\sum_q \Delta 
\<M_{\{\vec{j},q\},\{q,\vec{k}\}}\> = 0$ \footnote{We note that this does not imply that the individual terms 
of the sum vanish.}. Besides, if we consider, for instance, the case $M=2$, then we see that the right-hand 
side of Eq. \eqref{eq:hierCons} accounts only for single-particle contributions of two-particle 
clusters. 

In order to find a definition for $\<M_{\vec{j},\vec{k}}\>_{<M}$ such that Eq. \eqref{eq:hierCons} is 
fulfilled, let us first investigate, as an example, the case where $|\Psi\rangle$ is a general mean-field 
state, which is defined as a single permanent.
More precisely, given some single-particle orbitals $\chi_{j}(x)$ with the associated creation 
operators $\ald{j}$, a mean-field state is defined as a single permanent like $|\mt{MF}_\vec{k}\> = \prod_i^N 
\ald{k_i} \ket{\mt{vac}}$ with $\ket{\mt{vac}}$ being the vacuum and $ \vec{k} = \{ k_1,...,k_N\}$ (for the 
commonly known GP state $k_i=k_1 \, \forall i$). For such a state the two-particle clusters can be written as
\begin{align}
  \< \ad{k}  \ad{q} \a{q'} \a{k'} \>_\mt{MF} =&  \< \ad{k}  \a{q'} \>_\mt{MF} \< \ad{q}  \a{k'} \>_\mt{MF} 
\nonumber \\
&+ \< \ad{k} \a{k'}\>_\mt{MF} \< \ad{q}  \a{q'} \>_\mt{MF}  \nonumber \\
& + \Delta_\mt{B}^\mt{MF}(k,q,q',k') \label{eq:meanfield}
\end{align}
with the \textit{bosonic correlations} \footnote{The appearance of $\Delta_\mt{B}^\mt{MF}$ is 
due to the bosonic symmetry and the possibility for bosons to have an occupancy $n_j>1$.} given by
\begin{align}
 \Delta_\mt{B}^\mt{MF}(k,q,q',k') =& - \sum_j \bra{\chi_j}\ketO{\phi_k} \bra{\chi_j} \ketO{\phi_q}   
\nonumber \\
& \bra{\phi_{k'}}\ketO{\chi_j}  \bra{\phi_{q'}} \ketO{\chi_j} n_j(n_j+1). \label{eq:TBBmeanfield}
\end{align}
Here $n_j$ is the number of bosons in the single-particle state $\chi_{j}(x)$ defined via $n_j = \sum_i 
\delta_{k_i,j}$ ($\delta_{i,j}$ is 
the Kronecker-Delta). It is easy to verify that for such a mean-field state Eq. \eqref{eq:hierCons} is 
fulfilled if we set $\< \ad{k}  \ad{q} \a{q'} \a{k'} \>_{<M} = \< \ad{k}  \ad{q} \a{q'} \a{k'} \>_\mt{MF}$.
With this choice, the term $\Delta \<2_{\{k,q\},\{q',k'\}}\>$  is zero in a mean-field state per definition.
In this spirit, the terms $\Delta \langle M_{\vec{j},\vec{k}} \rangle$ can be understood as the 
$M$\textit{-particle  correlations}.


Now for a general many-body quantum state we replace in Eq. \eqref{eq:TBBmeanfield} 
the mean-field occupations $n_j$ and the mean-field basis functions $\chi_{j}(x)$ by the natural populations 
$\lambda_j$ and the natural orbitals $\Phi_j(x)$, respectively, yielding the following analogue expression:
\begin{align}
 \Delta_\mt{B}(k,q,q',k') =& - \sum_j \bra{\Phi_j}\ketO{\phi_k} \bra{\Phi_j} \ketO{\phi_q}   
\nonumber \\
& \bra{\phi_{k'}}\ketO{\Phi_j}  \bra{\phi_{q'}} \ketO{\Phi_j} \lambda_j(\lambda_j+1). \label{eq:TBbcorr}
\end{align}
Note that in general $\lambda_j$ is not an integer. Hence, we define the single-particle 
contributions of a two-particle cluster as
\begin{align}
  \< \ad{k}  \ad{q} \a{q'} \a{k'} \>_{<M} :=&  \< \ad{k}  \a{q'} \> \< \ad{q}  \a{k'} \> + \< 
\ad{k} \a{k'}\>\< \ad{q}  \a{q'} \>  \nonumber \\
& + \Delta_\mt{B}(k,q,q',k'), \label{eq:singPart}
\end{align}
whereas the two-particle correlations, also called \textit{doublets}, are defined as
\begin{align}
g_{kqq'k'} \equiv& \Delta \< \ad{k}  \ad{q} \a{q'} \a{k'} \> \nonumber \\
:=& \< \ad{k}  \ad{q} \a{q'} \a{k'} \> \nonumber \\
& - \< \ad{k}  \a{q'} \> \< \ad{q}  \a{k'} \> - \< \ad{k} \a{k'}\>\< \ad{q}  \a{q'} \>  \nonumber \\
& - \Delta_\mt{B}(k,q,q',k'). \label{eq:corrfac}
\end{align}

Here some considerations are in order. Our choice for the non-correlated part of the two-particle 
cluster [Eq. \eqref{eq:singPart}] indeed fulfills the condition \eqref{eq:hierCons} which 
justifies the above replacement of the mean-field occupations and basis functions by the natural populations 
and natural orbitals, respectively.  Although we focused here on the two-particle clusters, we 
note that expressions like Eq. \eqref{eq:corrfac} can be obtained for any $M$-particle clusters, too.
However, for the present study the singlets and doublets are sufficient to understand the dynamics of the
system.
Further, we would like to stress that the decomposition into singlets and $M$-particle correlations does not 
 correspond to a separation into mean-field and beyond mean-field contributions. But even if the 
condition \eqref{eq:hierCons} only separates 
the clusters into singlets, doublets, three-particle correlations, etc., it enables us to identify genuine
correlations of any many-body quantum state which are beyond mean-field. 
Indeed, in the limit of a mean-field state Eq. \eqref{eq:TBbcorr} boils down to Eq. \eqref{eq:TBBmeanfield}, 
and therefore all correlated parts $\Delta \langle 2_{\{k,q\},\{q',k'\}} \rangle = 0$ vanish. This would not 
be possible with the traditional cluster-expansion approach.

Finally, we would like to highlight that one could instead search for the best mean-field state 
\cite{Cederbaum2003} and then separate the $M$-particle clusters into mean-field part and contributions beyond 
that. It turns out, however, that such a choice does not satisfy Eq. \eqref{eq:hierCons}.
This shows that a mean-field approximation of a real many-body state does not necessarily result in the 
exact single-particle properties of the system.

\subsubsection{Singlet Dynamics}

Even though we are able to derive the time evolution of every $M$-particle cluster from our MCTDHB 
solutions, it is worth to investigate the equations of motion of the clusters since they make it 
possible to understand the coupling between the singlets themselves and between singlets and doublets. 
The dynamics of the $M$-particle clusters can be derived from the Heisenberg equation of motion and the 
above outlined definitions. For the analysis of the dynamics in Sec. \ref{sec:DynEvol}, we focus onto the 
equations of motion of the singlets. One can easily show that the time evolution of the coherences is given by
\begin{align}
 i\hbar \frac{\d}{\d t} p_{ij} =& 
\left( \tilde{\epsilon}_{j} - \tilde{\epsilon}_{i}^* \right)   p_{ij}  +\Sigma_{ji}(f_i-f_j) \nonumber \\
&+ \sum_{q\neq i,j} \left[ \Sigma_{jq} p_{iq} - \Sigma^*_{iq}p_{qj} \right] \nonumber \\
&  + \Gamma_{ij}^\mt{B} + \Gamma_{ij} \label{eq:pdyn}
\end{align}
while the one of the populations is governed by 
\begin{align}
 \hbar \frac{\d}{\d t} f_{i} =& 
2 \sum_{q\neq i} \Im \left[ \Sigma_{iq} p_{iq} \right] 
 + \Im \left[  \Gamma_{ii}^\mt{B} + \Gamma_{ii} \right].  \label{eq:fdyn}
\end{align}
Here we used the definition of Eq. \eqref{eq:corrfac} for the two-particle cluster that appears in the 
corresponding Heisenberg equation, because of the atom-atom interaction and the asterisk for the 
complex conjugation.  Besides, in the above equations, we have 
introduced the renormalized single-particle energies $ \tilde{\epsilon}_{i} = \epsilon_i 
 + \Sigma_{ii}$, the singlet couplings $ \Sigma_{ij} = 2\sum_{kq} V_{kijq} \langle \ad{k} \a{q} 
\rangle$, the coupling to the bosonic correlations
\begin{equation}
 \Gamma_{ij}^\mt{B} =  \sum_{kk'q}\left[   V_{kjqk'}  \Delta_\mt{B}(i,k,q,k') - V_{kqik'}   
\Delta_\mt{B}(q,k,j,k')  \right]  ,
\end{equation}
and to the doublets
\begin{equation}\label{eq:gamma_ij}
 \Gamma_{ij} =  \sum_{kk'q}\left[   V_{kjqk'} g_{ikqk'} - V_{kqik'} g_{qkjk'} \right]  ,
\end{equation}
and the interaction matrix elements
\begin{equation}
 V_{ijj'i'} = g\int \phi_i^*(x)\phi_j^*(x) \phi_{j'}(x)\phi_{i'}(x) \d{x}.
\end{equation}
Note that the equations of motion for the singlets form a system of non-linear 
differential equations, since the mean-field couplings are also defined by the singlets themselves.
Furthermore, they are coupled to the doublets (via $\Gamma_{ij}$) which can be interpreted as a 
source term (or inhomogeneity) for the singlets. On the other hand, the dynamics 
of the doublets depends on the singlets as well as on the three-particle correlations $\Delta \langle 3 
\rangle$, which renders the coupling time-dependent. 
We would like to emphasize that the coupling to $\Delta \langle 3 \rangle$ is a manifestation of the 
BBGKY-hierarchy.

For the truncation of the hierarchy at the singlet level, one has to ensure that the contribution of the 
doublets to the singlet dynamics remains very small during the time interval of interest, which would imply 
that one can set $\Gamma_{ij} = 0$.
In addition, we remind here again that in order to obtain a closed singlet theory, the 
coupling to the bosonic correlations has to be included which is the price to pay for our choice of 
factorization.
In the following, we show how a consistent and closed singlet theory can be accomplished. At first, by means 
of Eqs. \eqref{eq:hierCons} and \eqref{eq:singPart}, we obtain an exact relation among the bosonic 
correlations and the singlets of the system:
\begin{equation}\label{eq:consistency}
 \sum_q \Delta_\mt{B}(k,q,q,k') = -\sum_q \left[ \< \ad{k}  \a{k'} \>  \delta_{qk} +                
\<  \ad{k}  \a{q} \> \< \ad{q}  \a{k'} \> \right]. 
\end{equation}
Then, by using this relation and by noticing that 
$\Delta_\mt{B}(i,k,q,k') \neq 0$ only if all indices are even or if all are odd [see 
Eq. \eqref{eq:TBbcorr} and note that the natural orbitals have a defined parity in our setup] we can 
approximate 
$\Gamma_{ij}^\mt{B}$ as
\begin{align}
 \Gamma_{ij}^\mt{B} \approx&  \sum_q \Biggl[ V_{qqqi}\left(\<\ad{q}\a{j}\> +\sum_k 
\<\ad{q}\a{k}\>\<\ad{k}\a{j}\>\right) \nonumber\\
 &- V_{jqqq} \left( \<\ad{i}\a{q}\>  + \sum_k \<\ad{i}\a{k}\>\<\ad{k}\a{q}\> \right)    \Biggr].
\label{eq:TBbcorrApprox}
\end{align}
With this expression for the coupling to the bosonic correlations, the equations of motion of the singlets 
can be completely decoupled from higher than single-particle clusters such that a consistent and 
closed singlet theory is obtained. In the following, we will see the 
power of such a singlet theory in the analysis of the complicated many-body dynamics, especially in 
combination with the MCTDHB method.

\section{Dynamical Evolution}\label{sec:DynEvol}

We investigate now the dynamics induced by the instantaneous creation of an ion in an atomic 
cloud. First, we analyze the one-body 
density (matrix) and the components of the energy as well as the many-body excitation spectrum.
The observations are then discussed and analyzed in detail in terms of our singlet-doublet theory developed 
above.

\subsection{Observations}

\begin{figure}
 \centering
 \includegraphics[width=\linewidth]{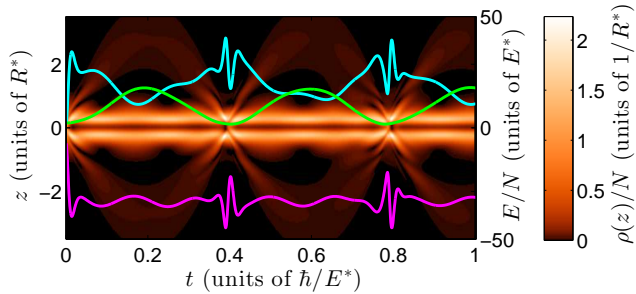}
 \caption{ Time evolution of the one-particle density $\rho(z,t)$ and the components of the total energy per 
particle for a system with $N=2$. The green, cyan, and magenta lines represent the trapping,  kinetic, and 
ionic energy per particle, respectively.}
\label{fig:dens_Energ}
\end{figure}

In Fig. \ref{fig:dens_Energ}, we show exemplarily the temporal evolution of the one-body 
density $\rho(z,t) = \rho(z,z,t)$ of the atomic cloud for $g=2 E^* R^*$ and $N=2$. 
In order to give a feeling of the actual time scales involved, we note that for 
$^{87}\mt{Rb}$ we have  $\hbar/E^* \approx 0.39\, \milli\second$, while for $^7\mt{Li}$ it corresponds to 
$\hbar/E^* \approx 0.001\, \milli\second$.
For short times, the suddenly created ion 
captures very quickly most of the atomic cloud in its bound states. The remaining atomic density fraction is 
emitted as a beam into the outer region of the harmonic trap. 
Consequently, this fraction is decelerated ($t<0.2\,  \hbar/E^*$) and back reflected ($0.2\, \hbar/E^* 
< t < 0.4\,  \hbar/E^*$) by the harmonic confinement. 
Subsequently, this sequence repeats with approximatively constant frequency. Furthermore, a second faster
oscillation in the density fraction captured within the ionic potential is visible (see the holes in the 
density plot at $z \approx \pm R^*/2$). Additionally, we show in Fig. \ref{fig:dens_Energ} the components of 
the energy per particle. 
The trapping energy (green line) perfectly oscillates with a single frequency which coincides with the 
oscillation frequency of the outer density fraction. In contrast, the ionic energy (magenta line), which 
represents the expectation value of the ionic potential \eqref{eq:modelPotpaper}, 
oscillates with a higher frequency matching the 
inner density oscillation. On top of this oscillation, we observe a short pulse when the outer fraction 
``crashes'' into the inner density part. Note that these events are not visible in the trapping energy since the harmonic trap energy is negligible in the vicinity of the trap center. The kinetic energy (cyan line) 
can be understood as the negativ sum of 
trapping and ionic energy, since the interaction energy (not shown) is comparably small and the total energy
is conserved during the dynamics. Hereafter, we term 
the oscillation of the outer fraction of the atomic cloud \textit{harmonic} and the inner fraction 
 \textit{ionic} oscillation. Note that the above observations are qualitatively independent of the atom 
number $N$ such that Fig. \ref{fig:dens_Energ} is representative also for larger $N$.

\begin{figure}
 \centering
 \includegraphics[width=\linewidth]{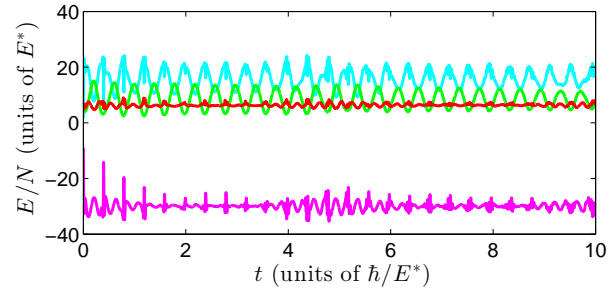}
 \caption{ Time evolution of the various  components of the energy per particle for $N=10$. The red, green, 
cyan, and magenta lines represent the interaction, trapping,  kinetic, and ionic energy per 
particle, respectively.}
\label{fig:energ}
\end{figure}

After multiple oscillation periods (see Fig. \ref{fig:energ}), we observe that the ionic 
energy (magenta line), and therefore the ionic oscillation, exhibits a clear collapse and revival behavior on 
this long time scale. On the other hand, the trapping energy (green line), and thus the harmonic 
oscillation, becomes only slightly damped. While for $N=2$ these two aspects of the long-time behavior can be 
barely seen, it becomes strongly pronounced for larger particle numbers.

\begin{figure*}
 \centering
 \includegraphics[width=\linewidth]{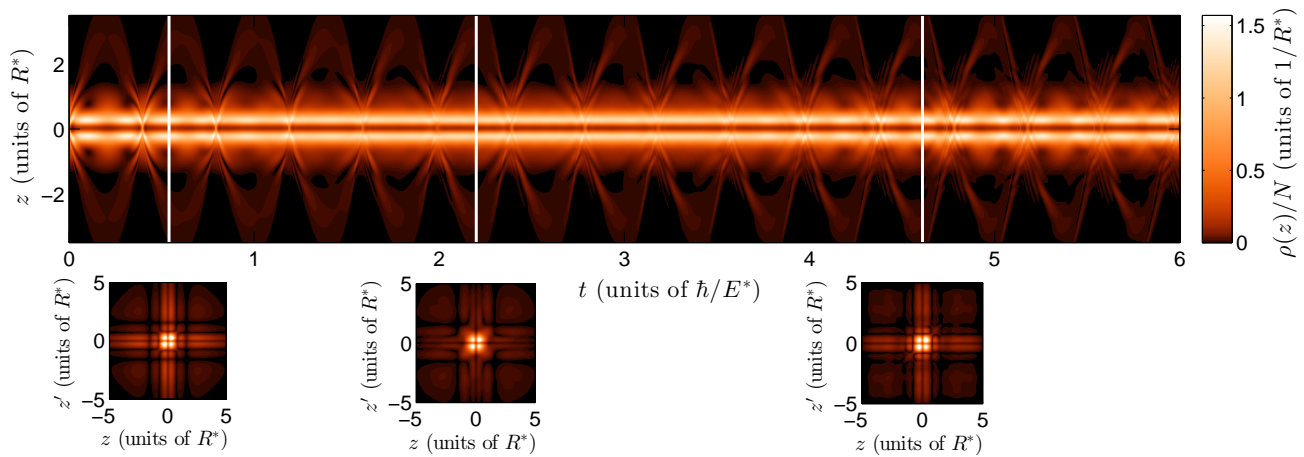}
 \caption{(Top panel) Time evolution of the one-particle density $\rho(z,t)$ for $N=10$. 
Snapshots of the one-particle reduced density
 matrix $\rho(z,z',t)$ (lower panels) at times $t= 0.54\, \hbar/E^*$ (left panel), $t= 2.20\, \hbar/E^*$ 
(middle panel), and $t= 4.61\, \hbar/E^*$ (right panel) which are indicated in the main top panel as white 
vertical lines ($\hbar/E^* = 0.39\, \milli\second$ for $^{87}\mt{Rb}$ atoms).}
\label{fig:dens}
\end{figure*}

The collapse and revival behavior can also be observed in the long time dynamics of the atomic density. 
Figure \ref{fig:dens} shows the disappearance of the ionic oscillation during the time interval $[1.5,3.5] 
\hbar/E^*$ and its recurrence around 
$t \approx  4.0\, \hbar/E^*$. 
This effect seems to be directly connected to the loss and regain of spatial coherence between the inner 
and outer density fraction which can be observed in the snapshots of the one-particle density matrix at times 
$t= 0.54\, \hbar/E^*$ (left panel), $t= 2.20\, \hbar/E^*$ (middle panel), and $t=  4.61\, \hbar/E^*$ (right 
panel) in Fig. \ref{fig:dens}. Below, we will understand the relation between the ionic oscillation 
and the spatial coherence in detail through the singlet-doublet analysis. 

\begin{figure}
 \centering
 \includegraphics[width=\linewidth]{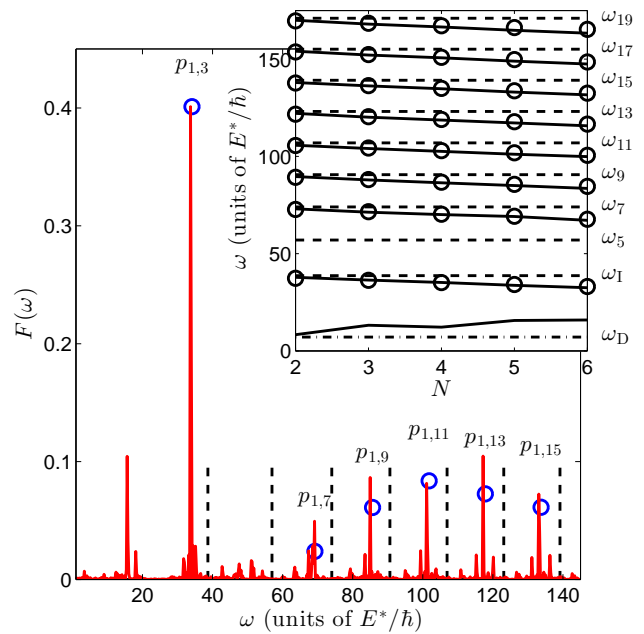}
 \caption{ (Main panel) Excitation spectrum for $N=5$ and $g=2E^*R^*$. The red line represents 
the actual Fourier spectrum  of the fidelity $F(t)$. The blue circles mark the single-particle energies (see 
text). We indicate the corresponding dominant singlet $p_{kq}$ at each peak by the label $p_{i,j}$ for the 
sake of better readability. The dashed vertical 
lines illustrate the non-interacting limit, thus correspond to $g=0$. Note that $p_{1,5}$ is hard to 
identify due to its small amplitude.
(Inset) Prominent resonances of the spectrum $F(\omega)$ in dependence of $N$. The black solid lines are 
 extracted from the Fourier transform of the fidelity $F(t)$. The circles indicate the resonance positions 
obtained from  our singlet theory (see text). Further, the excitation energies  for  the non-interacting 
case (g=0) $\hbar\omega_i = \epsilon_i - \epsilon_1$ (dashed lines),  $\hbar\omega_\mt{I} = \epsilon_3 - 
\epsilon_1$ (lowest dashed line), and  $\hbar\omega_\mt{D}=2(\epsilon_2-\epsilon_1)$ 
(dashed dotted line) are shown.}
\label{fig:spectrum}
\end{figure}

Let us now discuss the frequencies that are involved in the 
dynamics. To this aim, we investigate the fidelity defined by the overlap of the many-body wavefunction at 
time $t_0=0$ with the one at time $t$ \cite{Zanardi2006}:
\begin{equation}\label{eq:fidelity}
  F(t) = |\bra{\Psi(0)}\ketO{\Psi(t)}|^2.
\end{equation}
Since this fidelity $F(t)$ can be understood as the expectation value of the time-evolution operator, thus an 
$N$-body operator, its Fourier transform contains information of all involved excited
eigenstates of the interacting $N$ atom system. Due to the 
discrete nature of the spectrum and because in our numerical 
simulations the propagation time $T$ is finite, it turns out to be more efficient for the computation of the 
Fourier transform to use the compressed sensing (CS) method \cite{Donoho2006,Candes2006}.
With compressed sensing, one can indeed obtain a resolution in frequency space better than 
$\Delta\omega = \frac{2\pi}{T}$ \cite{Candes2006a}. To this end, we have used the matlab package ``SPGL1'' for 
compressed sensing from Ref. \cite{E.vandenBergandM.P.Friedlander2007}. The algorithms used in 
this package can be found in Refs. \cite{VandenBerg2008,VandenBerg2011}.
In Fig. \ref{fig:spectrum}, we show the Fourier transform $F(\omega)$ of the 
fidelity (red continuous line). Several prominent resonances become apparent. We observe 
one dominant mode at a frequency $\omega \approx 34 \, E^*/\hbar$. This corresponds to the ionic 
oscillation frequency. In contrast, the harmonic frequency can not directly be found in the spectrum.
Furthermore, we see that modes with high frequency are excited and seem to 
have an equidistant spacing. In addition to this, there is a low energy mode around $\omega 
\approx 15\, E^*/\hbar$, whose origin we will explain in the subsequent sections.

\subsection{Singlet Dynamics}

Let us start the analysis of the many-body spectrum in terms of the singlets. Therefore, we choose from 
here on the non-interacting single-particle functions $\phi^0_j(x)$ as the basis $\phi_j(x)$ in which the 
clusters are expressed.
In general, the identification of the modes corresponding to the observed resonances is a very complicated 
task. Nevertheless, we were able to identify the most important modes by linearizing the equation of motion 
for the singlets [see Eqs. \eqref{eq:pdyn} and \eqref{eq:fdyn}, and App. \ref{app:singletlin}].
The obtained energies together with the ``exact'' spectrum belonging to the Fourier transform of the fidelity 
[see Eq. \eqref{eq:fidelity}] are shown in Fig. \ref{fig:spectrum} (blue circles).
One observes a very good agreement of the obtained peak positions  with the 
Fourier spectrum of $F(t)$ (red continuous line). On the other hand, the relative heights of the peaks can be 
only obtained approximatively by our singlet theory (see App. \ref{app:singletlin}) showing qualitative 
agreement with the 
Fourier spectrum. 
Further, we can identify which singlet has the dominant contribution to a certain 
resonance (see App. \ref{app:singletlin} for details). 
This dominant singlet is indicated in Fig. \ref{fig:spectrum} at the corresponding peak.

We find that the most dominant frequency corresponds to $p_{13} = \< \ad{1}  \a{3} \>$ excitations. These are 
coherences between the number states  $|N,0,0,...\>$ and $|N-1,0,1,0,...\>$ (in the basis of the 
non-interacting single-particle states $\phi^0_j(z)$)
which oscillates with the frequency 
$\omega_\mt{I} = (\epsilon_3 - \epsilon_1)/\hbar = 38.7\, E^*/\hbar$ for $g=0$. Thus, the ionic 
oscillation 
corresponds to an oscillation between a bound state and the first trap state (see also the oblique magenta 
arrows in Fig. \ref{fig:potLevels}). Hence, it connects the 
inner part of the atomic cloud with the outer fraction establishing spatial coherence between them as we
have seen in Fig. \ref{fig:dens}. Moreover, 
this mode induces, as we will see in the following, population transfer between the inner and the outer 
atomic fraction.
On the other hand, the harmonic oscillation can not be attributed to a single resonance, because no resonance 
shows up at its frequency. Nevertheless, we 
can understand its origin. We note that singlets $\< \ad{1}  \a{i} \>$ with $i=5,7,9,...$ to very 
high $i$ are excited and therefore present in $F(\omega)$. They correspond to oscillations between the 
numberstate $|N,0,0,...\>$ and the states 
with $|N-1,0,...,0,1,0,...\>$, where a single particle is excited to the $i$th state and thus they oscillate, 
for $g=0$, with frequencies $\omega_i = (\epsilon_i - \epsilon_1)/\hbar$ with $i=5,7,9,..$ (note that 
excitations into states with $i$ even are symmetry forbidden). These frequencies are shown in Figs. 
\ref{fig:spectrum} as dashed lines. Since the difference between two neighboring
resonances is approximatively constant $ (\omega_{i+2} -\omega_i) \approx 2\omega_0$ due to the 
equidistant spacing of the energy levels in the harmonic trap, all of these singlets are in phase again after 
a time $T \approx \pi/ \omega_0 \approx 0.4 \hbar/E^*$.
Although this approximation does not work perfectly at all times, because of the presence of the ionic 
potential, the harmonic oscillation is visible for the entire simulation time due to the contribution of 
high energy modes which are only slightly affected by the presence of the ion. Nevertheless, the damping of 
the harmonic oscillation, visible in the trap energy (see also Fig. \ref{fig:energ}, green line), can be 
attributed to the induced slight dephasing.

In the inset of Fig. \ref{fig:spectrum}, the resonance positions in dependence of the particle number 
$N$ are shown. We observe that the peaks which can be associated to a coherence $p_{1j}$ shift to smaller 
energies for growing $N$. Our singlet theory 
is perfectly able to capture this behavior. Therefore, the shift of resonances can be understood as a 
mean-field phenomenon which effectively renormalizes the singlet resonances. Importantly, we 
numerically find that the property of equidistant spacing between the 
single-particle modes seems to be nearly untouched by the interaction and, as a consequence, the harmonic 
oscillation is essentially unaffected. 

The singlet theory explains most of the modes found in the Fourier spectrum $F(\omega)$. The low energy 
resonance at $\omega \approx 15.5\, E^*/\hbar$, however, does not appear in the singlet spectrum. Moreover, a 
closer inspection of the inset of Fig. \ref{fig:spectrum} shows that this mode has the tendency of an 
increasing energy as 
the number of bosons increases, which is the opposite behavior of the other resonances. We show in the 
following that this resonance can be attributed to correlations which are dynamically build-up during
the temporal evolution. To do so we have first to understand the coupling between the singlets and the 
doublets.

\begin{figure*}
 \centering
 \includegraphics[width=\linewidth]{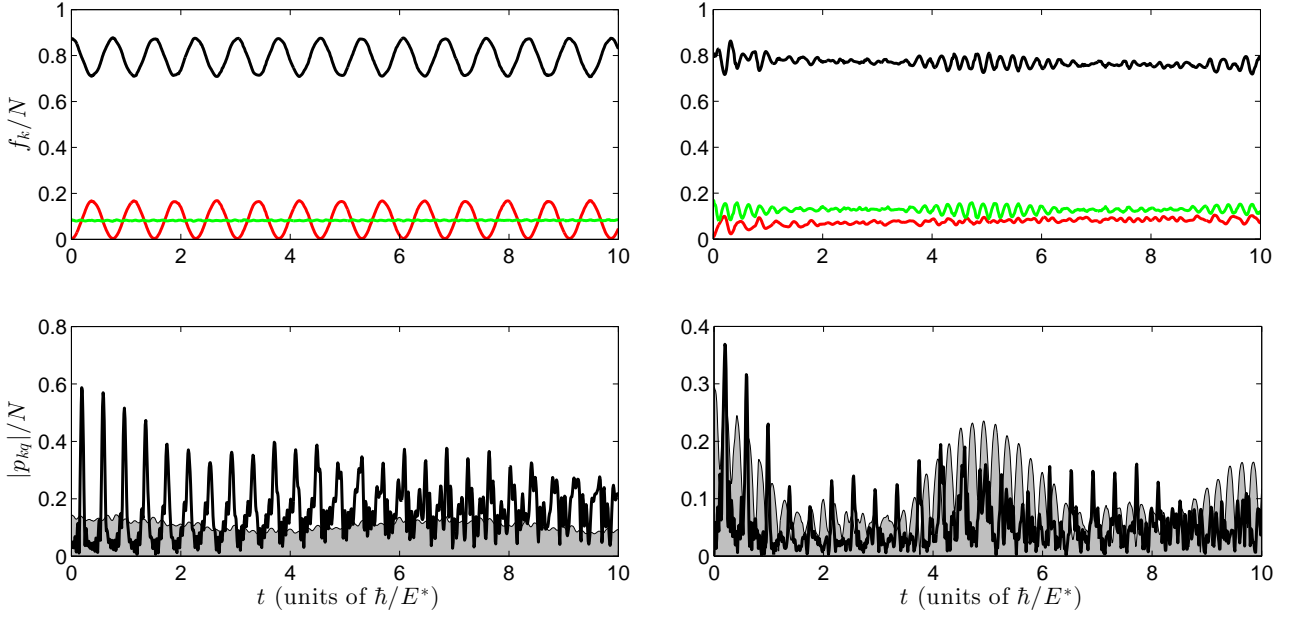}
 \caption{Time evolution of the singlets for $N=2$ (left panels) and $N=10$ (right panels). In the upper 
panels, the occupations $f_k$ are shown: $f_1$ (black line), $f_2$ (red line), and $f_\mt{T}$ (green line).
Further, the absolute values of the most important coherences $p_{kq}$ are plotted in the lower panel: 
$p_{1\mt{T}}$ as a black line, and $p_{13}$ as a gray shaded area (for better visibility).}
\label{fig:singletDyn}
\end{figure*}

Towards that end, we proceed further with the analysis of the dynamical evolution of the singlets. 
In Fig. \ref{fig:singletDyn}, the coherences 
$p_{13}$, $p_{1\mt{T}} = \sum_{k>3}p_{1k}$, and the occupations $f_1$, $f_2$, $f_\mt{T} = \sum_{k>3} f_k$ 
are shown. 
As stated beforehand, the coherence $p_{13}$ oscillates with the frequency of the ionic oscillation. From the 
time evolution of the ionic energy (see Fig. \ref{fig:energ}), one expects in addition the collapse and 
revival behavior. Here we see that for $N=2$ (lower left panel), the absolute value of $p_{13}$ is 
nearly constant, while collapse and revival become clearly visible for $N=10$ (lower right panel).
The coherence between the first bound and the trap states, 
$p_{1\mt{T}}$, shows the aforementioned rephasing behavior in the distinct peaks with $T \approx \pi/ 
\omega_0 \approx 0.4 \hbar/E^*$ periodicity. The damping is also visible here for $N=10$, but, importantly, 
the peaks are still 
very pronounced enabling the harmonic oscillation to be unaffected.
Turning our attention to the dynamics of the occupations (Fig. \ref{fig:singletDyn}, upper panels), we can 
identify that the population of the bound 
states, thus the ion population $f_\mt{I} = f_1 + f_2$, is about $80 - 90 \%$, and therefore only about $10\% 
-20\%$ of the particles are emitted into the trap by the ``ionization process''. Furthermore, a population 
transfer between the two bound states occurs (upper left panel). For larger $N$ 
(upper right panel), also transfer to and from the trap population (green line)
becomes visible with a rate equal to the frequency of the ionic oscillation.

An inspection of Eq. \eqref{eq:fdyn} reveals that the population transfer 
between the bound state $\phi^0_1$ and the trap states is mediated by coherences, primarily by $p_{13}$,  
because it has the highest contribution in the Fourier spectrum $F(\omega)$ (see also Fig. 
\ref{fig:spectrum}), explaining its 
oscillation with the frequency of the ionic oscillation. In contrast, the dynamical evolution 
of $f_2$ has to be induced by the correlations $\Gamma_{22}$, since all coherences in question vanish.
Hence, the population transfer from the state $\phi^0_{1}$ to the state $\phi^0_{2}$ would not take place in 
a mean-field scenario, thus it is 
a clear signature of a genuine many-body effect. We also note that 
the frequency of this population transfer is $\omega \approx 8.2\, E^*/\hbar$ for $N=2$ which 
matches very well to the position of the additional low energy peak seen in the inset of Fig. 
\ref{fig:spectrum}. 
Further, the larger $N$ is, the faster the transfer process between the first bound and the trapped states
happens which fits to the shift of the low energy resonance to larger energies visible in Fig. 
\ref{fig:spectrum}.
We therefore show, in the next section, which doublets play an important role for this oscillation and to
which process/mode it corresponds. Further, we will explain the origin of the collapse and revival of the 
ionic oscillation.

\subsection{Doublet Dynamics}

\begin{figure*}
 \centering
 \includegraphics[width=\linewidth]{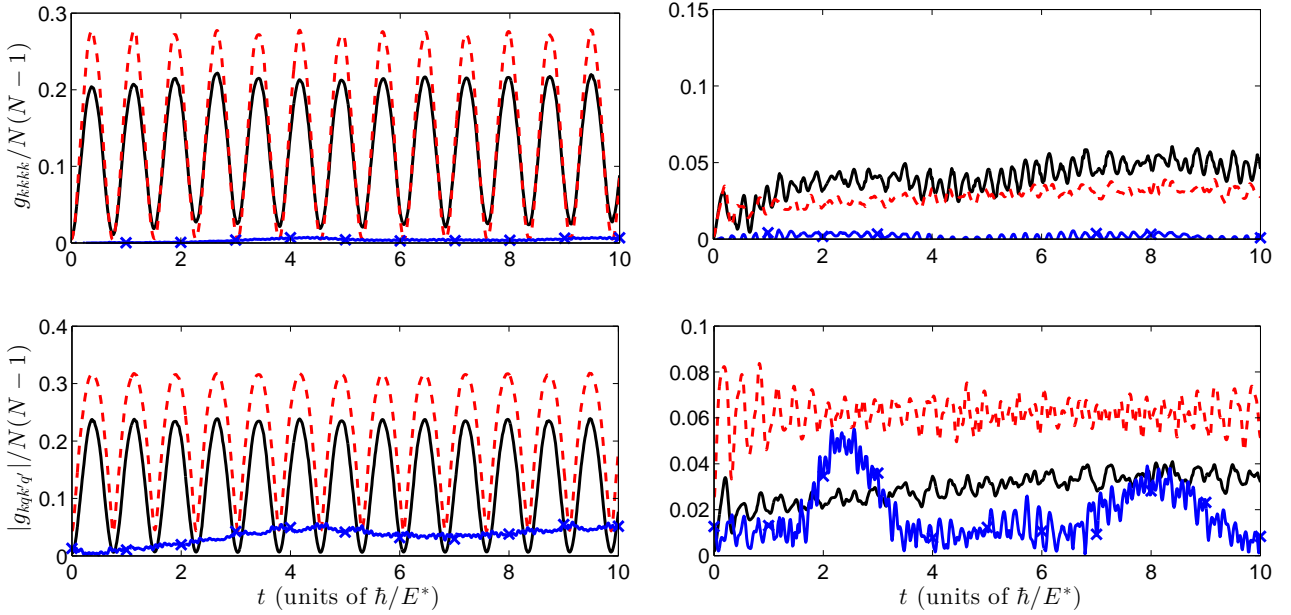}
 \caption{Time evolution of the most important doublets for $N=2$ (left panels) and $N=10$ (right panels). In 
the upper panels, the incoherent doublets $g_{1111}$ (black solid line), $g_{2222}$ (red dashed line), and 
$g_{3333}$ (blue line with crosses) are shown while the absolute values of coherent doublets $g_{1221}$ 
(black solid line), $g_{1122}$ (red dashed line), and $g_{1133}$ (blue line with crosses) are shown in the 
two lower panels.}
\label{fig:doubletDyn}
\end{figure*}

In order to understand the observed collapse and revival phenomena and the incoherent population transfer, we 
need to investigate the dynamical evolution of the doublets. For the sake of clarity, we classify the 
most important doublets (see definition in Sec. \ref{sec:defCluster}) into \textit{incoherent doublets} 
$g_{kkkk}$, 
$g_{kqkq}$, and \textit{coherent doublets}
 $g_{kkqq}$, $g_{kqkk}$. 
This separation is justified by the fact that the incoherent doublets are always real-valued quantities, and 
therefore do not 
contribute to the dynamics of the singlet occupations [see Eq. \eqref{eq:fdyn}]. 
Moreover, the doublets have the properties $g_{kqk'q'} = g_{k'q'kq}^*$ and $ g_{kqk'q'} = g_{kqq'k'} = 
g_{qkk'q'} = g_{qkq'k'}$. Therefore, we can identify, noting that only correlations involving 
states up to $k=3$ are contributing to the system dynamics, the most important classes of 
correlations: incoherent
$\{g_{1111},g_{2222},g_{3333},g_{1212},g_{1313} \}$ and coherent 
$\{g_{1122},g_{1133},g_{2233},g_{1211},g_{2122},g_{1311},g_{3133},g_{2322},g_{3233} \}$.
In Fig. \ref{fig:doubletDyn}, we show for $N=2$ (left panels) and $N=10$ (right panels) only those doublets 
of the incoherent (upper panels) and the coherent (lower panels) doublets which are considerably occupied. At 
first, we 
observe that, even though we start in an essentially uncorrelated state, very quickly a tremendous amount of 
correlations is created in the course of the temporal evolution. With the help of these correlations, we are 
now able to answer the remaining open questions concerning the dynamical evolution of the system.

In order to understand the population transfer between the states $\phi^0_{1}$ and $\phi^0_{2}$, we have to 
identify the correlations $g_{qkk'q'}$ contributing to $\Gamma_{11}$ and $\Gamma_{22}$. The coherent 
doublets $g_{1122}$ ($g_{2211}$) is the only correlation in question which can be understood by inspecting 
Eq. \eqref{eq:gamma_ij} and Fig. \ref{fig:doubletDyn}. Comparing Figs. \ref{fig:singletDyn} and 
\ref{fig:doubletDyn} (left panels), we 
see that they are build up and decay with the frequency of the population transfer between the state 
$\phi^0_{1}$ to the state $\phi^0_{2}$.
Since $g_{1122}$ and $g_{2211}$ which drive $\Gamma_{11}$ and $\Gamma_{22}$, respectively, are by $\pi$ out of 
phase ($g_{1122} = g_{2211}^*$), they induce the population transfer between those two bound states. Thus, 
we can understand these coherent doublets as mediators for the population transfer.
Further, we can now identify the low energy mode in the spectrum of Fig. \ref{fig:spectrum} as a coherent 
oscillation between the 
numberstates $|N,0,...\>$ and  $|N-2,2,0,...\>$ which corresponds to a two-particle excitation. In the 
non-interacting 
case, this mode would therefore oscillate with a frequency 
$\omega_\mt{D}=2(\epsilon_2-\epsilon_1)/\hbar
\approx 7.4\, E^*/\hbar$ indicated by a dashed dotted line in the inset of Fig. \ref{fig:spectrum}.
Here we can understand the following: First, this oscillation 
can not be seen in single-particle quantities like the one-body reduced density matrix. Second, this is also 
the reason why we could not explain the low energy peak by our singlet theory. Third, any mean-field 
approach would not be able to capture the associated dynamics. For example, in a GP theory the 
population transfer from $|N,0,...\>$ to $|N-2,2,0,...\>$ would be not possible, because in a 
general GP state $ (\sum_k c_k \ad{k})^N \ket{\mt{vac}}$ contributions from odd states are symmetry forbidden 
as long as only GP states with parity symmetry are considered.

Finally, we would like to discuss possible reasons for the collapse and revival behavior 
of the ionic oscillation occurring for larger  $N$. The 
impact of the doublets onto the singlet $p_{13}$ is contained in $\Gamma_{13}$. 
Even if many doublets contribute to $\Gamma_{13}$, we can identify the responsible doublet by inspecting 
Fig. \ref{fig:doubletDyn} (right panels). We observe that exactly at the times when the coherent doublets 
$g_{1133}$ are present, 
 the ionic oscillation is strongly suppressed whereas when $g_{1133}$ nearly vanishes the ionic oscillation 
reappears (compare Figs. \ref{fig:energ}, \ref{fig:dens}, and \ref{fig:singletDyn}). 
Therefore, the doublet $g_{1133}$ is 
the main source of loss of the coherence $p_{13}$. Consequently, this doublet is responsible for the loss of 
spatial coherence between the inner and the outer density fraction during the dynamics, as it can be seen in 
Fig. \ref{fig:dens}. 
Nevertheless, the build-up of $g_{1133}$ does not only act as a source of damping for the singlet coherences, 
since the coherences $p_{13}$ recur when the value of $g_{1133}$ is reduced again.
Further, we note that $g_{1133}$ ($g_{3311}$) is also present in $\Gamma_{11}$ ($\Gamma_{33}$) such that also 
the collapse and revival in the population transfer can be attributed to this coherent doublet.

\section{Convergence Analysis}\label{sec:conver}

In the following, we briefly discuss the quality of our data and 
to which extend advanced tools as MCTDHB in the description of the dynamical behavior of such quantum 
systems are indeed necessary. To this aim, we inspect the natural populations [see Eq. \eqref{eq:natorbs}] 
which provide an assessment of the degree of fragmentation of the system \cite{Penrose1956}.
Further, one can use them to systematically judge the convergence of our result to the 
exact solution of the many-body quantum system \cite{Beck2000}.
\begin{figure}
 \centering
 \includegraphics[width=\linewidth]{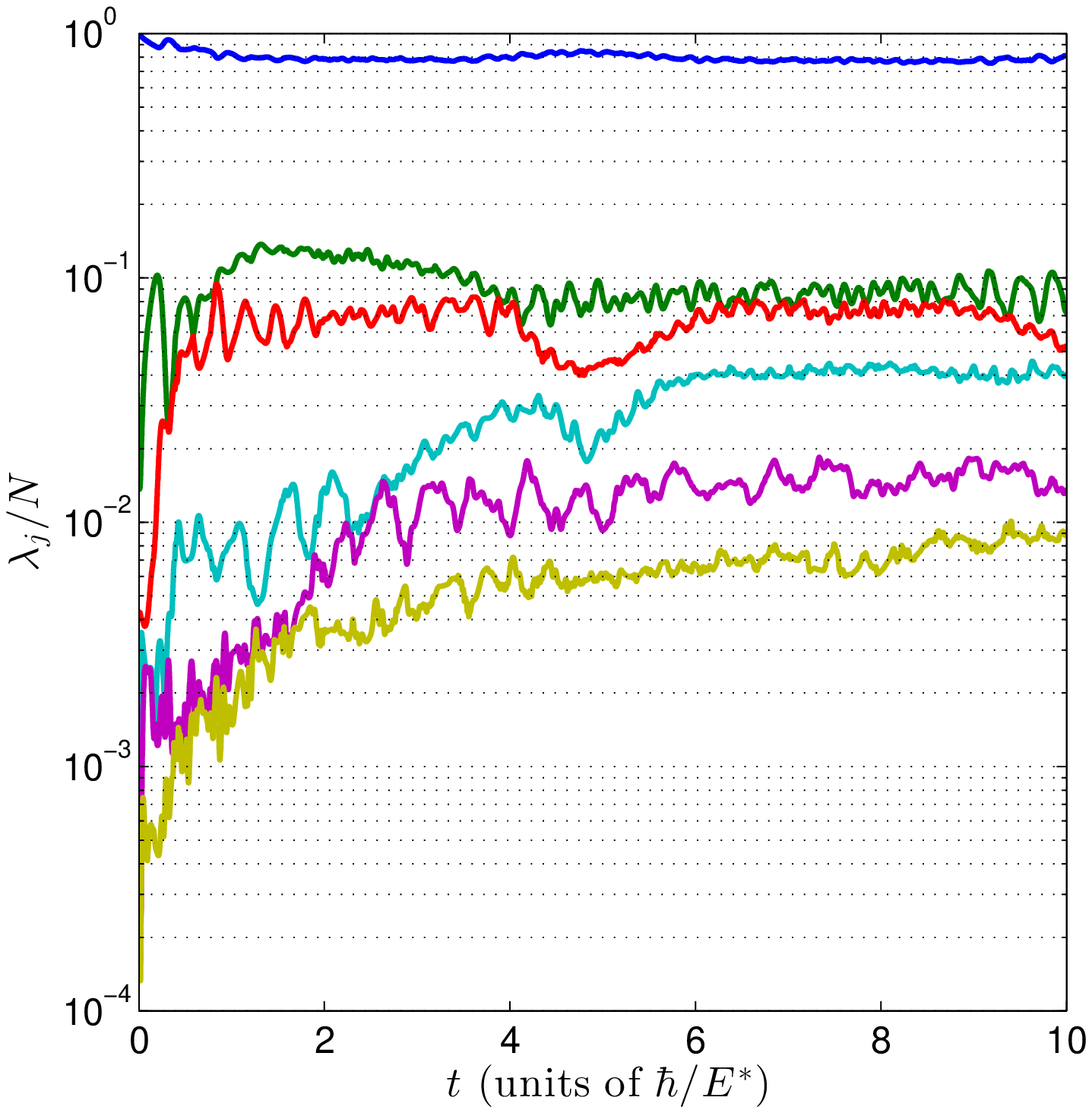}
 \caption{ Natural populations $\lambda_j(t)$ exemplarily for $N=10$ particles on a 
logarithmic scale. Note that the $\lambda_j$ are by definition ordered such that $\lambda_1 > \lambda_2 > ... 
> \lambda_m$ for all times.}
\label{fig:natpops}
\end{figure}
In Fig. \ref{fig:natpops}, the evolution of the natural populations exemplarily for $N=10$ 
particles is shown. We see that even if the initial state is nearly condensed, thus describable in the 
GP framework, depletion from the condensate state is created very quickly during the time evolution. 
Therefore, a GP description would be inaccurate and a multi-orbital description 
becomes unavoidable. Here we are using $m=6$ single-particle orbitals. The 
contribution of the lowest natural orbital (smallest natural population) stays below $1\%$
such that only natural orbitals with even smaller contribution are expected to be not taken into account by 
truncating the Hilbert space. One could further speculate that in such cases a multi-orbital mean-field 
description might  be sufficient. However, this is not the case and we stress that the 
non-steady behavior of the natural populations implies the necessity to go beyond a mean-field description 
\cite{Kronke2015}. Therefore, the utilization of MCTDHB is essential to 
capture the full correlated quantum dynamics of the system.
\section{Conclusions and Outlook}\label{sec:concl}

We have investigated the dynamics of a trapped cloud of $N$ interacting bosons after 
the sudden creation of a static ion. Thereby, most of the atoms are quickly captured in the 
bound states of the atom-ion potential, while the remaining atomic fraction is emitted into the trap. We were 
able to understand the subsequent dynamics and the many-body spectrum in detail by means of an underlying 
singlet-doublet theory.

Our atom-ion hybrid system exhibits, apart from the harmonic oscillation, that is, a density oscillation 
within the external harmonic confinement, an additional density oscillation which 
we were able to attribute to a coherent oscillation between a bound state and a trap state. This coherent 
oscillation results  in the spatial coherence between the inner and outer density fraction. It also allows 
for population transfer between bound and trap states 
such that the fraction of atoms in the bound states becomes time-dependent. In contrast to the harmonic 
oscillation, this ionic oscillation shows a particle number dependence.
Furthermore, we showed that, even though the atoms are weakly interacting, strong correlations are build-up 
during the temporal evolution. These correlations lead to population transfer between the 
bound states and show up as a strong doublet resonance in the many-body spectrum. Besides this, the 
correlations induce a periodic suppression of coherence between the inner and the outer fraction which 
gives rise to collapse and revival of the ionic oscillation on longer time scales.

The impact of the ion onto the bosonic cloud dynamics can be summarized as follows: Apart from the 
accumulation of atoms on both sides of the ion while depleting the cloud at the ion position,
the ion induces a coherent 
oscillation between the bound states and the 
trap states. The oscillation essentially arises because of the additional scales present in the system. 
Indeed, this can be easily seen in the non-interacting case ($g=0$) and taking the 
limit of small trapping frequency $\omega_0 \rightarrow 0$. In this case, the frequency of the ionic 
oscillation converges to the 
non-interacting single-particle energy $\epsilon_1$  of the lower bound state (see Sec. 
\ref{subsec:ModelofSys} for definition) 
which, thus, sets effectively a lower bound to the frequency of the ionic oscillation. Hence, although we 
have chosen a tight trap, which is numerically convenient to resolve the 
dynamics in a finite system, the ionic oscillation would be present with a frequency on the 
same order of magnitude for a weak confinement, too.

The present work can be viewed as a further step towards the simulation of the dynamics of the hybrid 
atom-ion system in the ultracold regime. 
The fact that quasi one-dimensional quantum Bose gases are routinely realized in several laboratories and 
given the recent advances in optical trapping of ions, puts the experimental realization of the hybrid 
system considered here within reach.
Furthermore, since the multilayer extension of our method (ML-MCTDHB) is especially designed for the 
simulation of mixtures, we plan to investigate the impact of the ionic motion onto the ground state and the 
dynamical evolution of the bosonic ensemble in the nearer future. This will be done by treating the ion 
quantum mechanically as well. Due to the attractive 
inter-particle interaction, these future studies can 
reveal the dynamical formation of molecular ions as predicted in Ref. \cite{Cote2002}. Furthermore, it would 
be of interest to study the energy transfer between ionic and atomic degrees of freedom in order to 
understand, for example, sympathetic cooling of the ion in the atomic cloud in more detail.

\section{Acknowledgements}
JS thanks Sven Kr\"onke, Lushuai Cao, and Valentin Bolsinger for many discussions.
This work has been financially supported by the excellence cluster 'The Hamburg Centre for
Ultrafast Imaging - Structure, Dynamics and Control of Matter at the Atomic Scale' of
the Deutsche Forschungsgemeinschaft.


\appendix

\section{Derivation of Clusters from Density Matrices}\label{sec:deriveClusters}

The $M$-particle clusters can be calculated via the $M$-particle reduced density matrices. Let us start with the 
time-dependent one-particle reduced density matrix which describes the spatial density and coherence of the 
system. We express it in terms of the natural orbitals $\Phi_j$ and the natural populations $\lambda_j$ as 
\begin{equation}
 \rho(x,x',t) = \sum_j \lambda_j(t) \Phi_j^*(x,t)\Phi_j(x',t).
\end{equation}
If we now expand the natural orbitals in an arbitrary single-particle basis $\phi_j(x)$
\begin{equation}
 \Phi_j(x,t) = \sum_k c_{jk}(t) \phi_k(x)
\end{equation}
with 
\begin{equation}
  c_{jk}(t) = \int \phi^*_k(x) \Phi_j(x,t) \d{x}
\end{equation}
and express also $\rho(x,x',t)$ in this basis
\begin{equation}
 \rho(x,x',t) = \sum_{k,q} \phi_k^*(x)\phi_q(x') \langle \ad{k} \a{q} \rangle,
\end{equation}
we can identify the single-particle cluster as
\begin{equation}
   \langle \ad{k} \a{q} \rangle = \sum_j c_{jk}^*(t) \lambda_j(t) c_{jq}(t).
\label{eq:deriveSinglets}
\end{equation}
Here $\ad{k}$ and $\ag{k}$ are the creation an annihilation operators of the single particle basis 
$\phi_k(x)$.
In the same manner, we can proceed in order to derive the two-particle clusters. 
By defining the expansion coefficients as
\begin{equation}
  d_{kq}^j(t) = \int \phi^*_k(x_1) \phi^*_q(x_2) \Phi_j(x_1,x_2,t) \d{x_1}\d{x_2},
\end{equation}
where now $\Phi_j(x_1,x_2,t)$ are the geminals, that is, the eigenfunctions of the two-body density matrix, 
we obtain the following expression for the two-particle clusters
\begin{equation}
   \langle \ad{k} \ad{q} \a{q'} \a{k'} \rangle = \sum_j d_{qk}^{j*}(t) \gamma_j(t) d_{k'q'}^j(t).
\label{eq:deriveDoublets}
\end{equation}

\section{Linearized Singlet Dynamics}\label{app:singletlin}

In order to derive the energy spectrum of the singlets, we need to linearize the equations of motion 
\eqref{eq:pdyn} and \eqref{eq:fdyn}. Even though such an approximation can not be able to describe the 
full dynamics, we should be able to predict the spectrum of the singlets 
rather accurately.
To this end, we write the singlets as
\begin{equation}
 \langle \ad{i} \a{j} \rangle(t) = \langle \ad{i} \a{j} \rangle_0 + \delta \langle \ad{i} \a{j} \rangle(t),
\end{equation}
where $\langle \ad{i} \a{j} \rangle_0$ is the time-independent mean value of $\langle \ad{i} \a{j} 
\rangle(t)$ and $\delta\langle \ad{i} \a{j} \rangle(t)$ is a small time-dependent fluctuation. By 
neglecting the terms quadratic in $\delta\langle \ad{i} \a{j} \rangle$, we get a system of linear 
differential equations for $\delta\langle \ad{i} \a{j} \rangle$. We obtain
\begin{align}
 i\hbar \frac{\d}{\d t} \delta \langle \ad{i} \a{j} \rangle =& 
\sum_{kq} \Big[ \tilde{\epsilon}_{jq}^0 \delta_{ki} - \tilde{\epsilon}_{ik}^{0*}\delta_{qj}  \nonumber \\
& + 2\sum_{k'}V_{kjk'q}\langle \ad{i} \a{k'} \rangle_0 \nonumber \\
&- 2\sum_{k'}V_{qk'ik}\langle \ad{k'} \a{j} \rangle_0 \nonumber \\
&+ (V_{kkki}  +\sum_{q'}V_{q'kki} \<\ad{q'}\a{k}\>_0 )\delta_{jq} \nonumber \\
&- (V_{jqqq}  +\sum_{q'}V_{jqqq'} \<\ad{q}\a{q'}\>_0 )\delta_{ik} \nonumber \\
&+ V_{kqqi} \<\ad{q}\a{j}\>_0 - V_{jkkq} \<\ad{i}\a{k}\>_0
\Big] \delta \langle \ad{k} \a{q} \rangle \nonumber \\
&+ \tilde{\Gamma}_{ij}  \label{eq:singledynLin}
\end{align}
where $\tilde{\epsilon}_{ij}^0 = \epsilon_i \delta_{ij}
 + 2\sum_{kq} V_{kijq} \langle \ad{k} \a{q} \rangle_0$  and $\Gamma_{ij}^{\mt{B}0}$ coincides with the 
definitions of  $\Gamma_{ij}^{\mt{B}}$, but with the mean values of the singlets inserted, 
whereas $\tilde{\Gamma}_{ij}$ is defined as
\begin{equation}
 \tilde{\Gamma}_{ij} = \Gamma_{ij} + \Gamma_{ij}^{\mt{B}0} + \sum_{q} \left[ \tilde{\epsilon}_{jq}^0 \langle 
\ad{i} \a{q} \rangle_0  
- \tilde{\epsilon}_{iq}^{0*} \langle \ad{q} \a{j} \rangle_0 \right].
\end{equation}
Here we used Eq. \eqref{eq:TBbcorrApprox} for the correlations $\Gamma_{ij}^{B}$. This is an important 
step because in case the doublets are negligible it decouples the singlets dynamics from any other 
time-dependent driving such that the many-body system can be described by singlets only.
Finally, Eq. \eqref{eq:singledynLin} can be rewritten in matrix form as
\begin{equation}
  i\hbar \frac{\d}{\d t} \vec{S}(t) = M \vec{S}(t) + \tilde{\vec{\Gamma}}(t).
\end{equation}
such that the homogeneous solution can be expanded in terms of the eigenvalues $\nu_n$ 
and the left(L) or right (R) eigenvectors $\vec{S}_n^\mt{L,R}$ of the matrix $M$ 
\begin{equation}\label{eq:matrix}
 M \vec{S}^\mt{R}_n = \nu_n \vec{S}^\mt{R}_n,
\end{equation}
and the imhomogeneous part can be found by means of the variation of the constant method leading to the full 
solution
\begin{equation}
 \vec{S}(t) = \sum_n \left[ c_n   + \int_0^t \frac{1}{i\hbar}  e^{i\nu_n 
t'/\hbar} \vec{S}^{\mt{L}*}_n \tilde{\vec{\Gamma}}(t') \d{t'}\right] \vec{S}^\mt{R}_n e^{-i\nu_n t/\hbar}  .
\end{equation}
The coefficients $c_n$ determine how strong the mode $\vec{S}^\mt{R}_n$ is excited and can be obtained by 
projecting the above solution onto the eigenvector basis, resulting in
\begin{equation}\label{eq:osciStrength}
 c_n = e^{i\nu_n t/\hbar} \vec{S}^{\mt{L}*}_n \vec{S}(t) - \int_0^t \frac{1}{i\hbar}  e^{i\nu_n 
t'/\hbar} \vec{S}^{\mt{L}*}_n \tilde{\vec{\Gamma}}(t') \d{t'}.
\end{equation}

In summary, we see that the linearized singlet equations provide us with the singlet resonances $\nu_n$ and 
the singlet modes $\vec{S}_n^\mt{R}$. We can derive these from the MCTDHB solution in the following way: 
Via Eq. \eqref{eq:deriveSinglets}, we can extract the dynamics of the singlets which we can use to derive the 
matrix $M$, and therefore the solution $\vec{S}(t)$. By diagonalizing $M$, we obtain the singlet energies and 
the dominant singlet in the eigenmodes (largest entry in the vector) shown in Fig. \ref{fig:spectrum}. 
Further, we can extract the oscillator strength $|c_n|^2$ by Eq. \eqref{eq:osciStrength} which we can use as 
a measure for the relative height of the peaks in the many-body Fourier spectrum of $F(t)$ just by scaling 
them to the global maximum of the spectrum. Note that, Eq. \eqref{eq:osciStrength} 
should be evaluated at small $t$ in order to allow the linear approximation.


\bibliography{library}


\end{document}